\documentclass[12pt,oneside,reqno]{article}
\usepackage{subcaption}
\usepackage{changepage}
\usepackage{amsfonts}
\usepackage{geometry}
\usepackage{amsthm}
\usepackage{amsmath}
\usepackage{amssymb}
\usepackage{amsfonts}
\usepackage{bbm}
\usepackage{verbatim}
\usepackage{threeparttable}
\usepackage{caption}
\newcommand{\captionnote}[1]{%
  {\captionsetup{justification=justified, font=footnotesize} \caption*{#1}}%
}

\usepackage{textcomp}
\usepackage{booktabs}
\usepackage{graphicx}
\usepackage{setspace}
\usepackage{multirow}
\usepackage{color}
\usepackage{relsize}
\usepackage{titling}
\usepackage[cmyk,x11names]{xcolor}
\usepackage[utf8x]{inputenc}
\usepackage{dsfont}
\usepackage{pgf,pgfplots,import}
\pgfplotsset{compat=newest}
\usepackage[normalem]{ulem}
\definecolor{accentcolor}{RGB}{50, 70, 150}
\usepackage{hyperref}
\hypersetup{
    pdfborderstyle={},
    colorlinks=true,
    linkcolor=accentcolor,
    urlcolor=accentcolor,
    citecolor=blue
}

\usepackage{natbib}

\usepackage[]{ifdraft}

\usepackage[font=normalsize,justification=centering]{caption}
\DeclareCaptionLabelFormat{captionf}{\textsc{#1~#2}}
\usepackage{lscape}
\usepackage{placeins}
\providecommand{\U}[1]{\protect\rule{.1in}{.1in}}
\title{Should I State or Should I Show? Aligning AI with Human Preferences\thanks{We thank Simon Angus, Joy Buchanan, Guillaume Fr\'echette, Yoram Halevy, Annie Liang, Daniela Scur, and Simon Wilkie, as well as seminar and conference participants at Deakin University, University of Technology Sydney, RMIT University, ESA Asia-Pacific 2026, BSE Summer Forum 2026, and SDM 2026 for their helpful comments and suggestions. We gratefully acknowledge funding from the Department of Economics Research Committee at Monash University. This project was reviewed and approved by the Monash University Human Research Ethics Committee under project number 50731.}}

\author{Keaton Ellis\thanks{Department of Economics, Monash University. Email: \href{mailto:keaton.ellis@monash.edu}{keaton.ellis@monash.edu}}\qquad  Wanying Huang\thanks{Department of Economics, Monash University. Email: \href{mailto:kate.huang@monash.edu}{kate.huang@monash.edu}}}

\begin{document}

\maketitle

\onehalfspacing

\begin{abstract}
The proliferation of AI agents introduces a new principal-agent problem which stems from human principals' difficulty in articulating preferences. We report results from an experiment focused on mitigating this problem via \textit{revealed} preferences from choice data. Compared to \textit{stated} preference from human-written prompts, individuals communicate their preferences more effectively through choices, with as few as two sufficing to match prompts' predictive value. This gap largely reflects subjects' difficulty in translating preferences into prompts and is largest among those exhibiting more Allais-type behavioral patterns. Subjects also misperceive the approaches’ relative performance, often choosing the less accurate one. 

\end{abstract}

 \newpage
\section{Introduction}\label{sec:introduction}

The emergence of agentic artificial intelligence (AI) has prompted widespread discussion of a future in which AI agents act autonomously on behalf of humans, with little or no direct oversight. Although views may differ on the exact form this AI-assisted future will take, there is a growing scholarly consensus that AI will fundamentally transform economic interactions and productivity (e.g., \citealp{brynjolfsson2025generative, acemoglu2025simple, gans2026optimal}). At the same time, the proliferation of AI agents introduces a new form of principal-agent problem in which misalignment may arise not because AI agents have preferences of their own, but because human principals cannot fully articulate their preferences.

As a simple example, consider a traveler, the principal, who tasks an AI agent with ``buying the cheapest flight from Melbourne, Australia to Washington, D.C.'' The agent may select an itinerary with an extremely long layover\textemdash an outcome the principal might implicitly wish to avoid but did not explicitly rule out in their instructions.
Consequently, the AI agent may make a decision that is irreversible and costly, e.g., the purchase of a cheaper but nonrefundable flight ticket. Indeed, as AI agents take on a more autonomous role in decision-making \citep{immorlica2024generative},  communicating preferences effectively to them becomes increasingly important. More generally, individuals may be unable to fully specify their preferences, and such incompleteness is likely to persist. Thus, it is a first-order challenge to identify channels that mitigate the resulting misalignment. 

To address this challenge, in this paper, we adopt a \textit{revealed preference} approach that uses human principals' past choices as a direct channel for communicating their preferences to an AI agent. We compare this approach with one based on \textit{stated preferences}, elicited through incentivized prompts that human principals write to guide the AI agent to make decisions on their behalf. The question we ask is: which approach 
is more effective for the human principals to communicate their preferences to an AI agent in the domain of choice under risk? 

The answer to the above question is \textit{a priori} unclear. On the one hand, stated preference information can, in principle, be arbitrarily precise, while revealed preference information typically cannot pin down a preference based on a limited number of observations \citep{varian1982nonparametric}. Thus, if human principals can precisely articulate their preferences, an AI agent given those preferences may outperform one given only past choices. On the other hand, human principals may struggle to articulate their preferences \citep{imas2025agentic, shahidi2025coasean}, so their choices may convey more information than their written prompts. 

We study which approach is more effective by conducting an incentivized online experiment that elicits both revealed and stated preference information. The experiment consists of two main parts. In Part I, subjects answer a series of 13 binary choice problems over lotteries with varying levels of difficulty. They are then asked to write a prompt stating their preferences to an AI agent that will later act on their behalf in a new series of similar lottery choice problems. In Part II, subjects make their own choices in these new problems. To evaluate the effectiveness of each approach, for each subject, we instantiate two AI agents using Anthropic's Claude Opus 4.5\textemdash a frontier large language model at the time of experiment. The first agent, \textbf{Data-AI}, is given all of the subject's 13 choices in Part I; the second agent, \textbf{Prompt-AI}, is given the subject's written prompt. We use the proportion of subjects' Part II choices that each AI agent correctly predicts as a measure of out-of-sample prediction accuracy. We refer to it as the \textit{match rate}. 

We first find that, on average, subjects communicate their preferences more effectively through their choices than through their written prompt. Data-AI correctly predicts 74\% of subjects’ Part II choices, compared with 71\% for Prompt-AI, a statistically significant advantage that is robust across question difficulty. Moreover, we find that much of this predictive value can be obtained from very few observations: providing Data-AI with as few as \textit{two} observed choices is sufficient to achieve the match rate of Prompt-AI. 

We next investigate why revealed preference information is more effective and find that its advantage is largely driven by subjects’ difficulty in articulating their preferences in written prompts. We instead generate prompts via an AI facing the same environment as subjects and then instantiate another AI agent given these prompts. The resulting AI agents are substantially more predictive, and their performance is on par with Data-AI. Additionally, as subjects' written prompts become more semantically similar to those generated by AI, the performance gap between Data-AI and Prompt-AI essentially closes. This suggests that there exist informative prompts that can be as effective as the choice data itself, yet subjects struggle to write them.

The aggregate advantage of the revealed preference approach, however, masks substantial heterogeneity across subjects. About half of the subjects communicate their preferences to an AI agent strictly more effectively through choices, one quarter more effectively through prompts, and one quarter equally effectively. When we examine how this heterogeneity correlates with subjects' observed characteristics, a key determinant is how often they exhibit Allais-type behavioral biases \citep{allais1953comportement}. We find that subjects who more frequently exhibit these biases are harder to predict from either information type, but the relationship is asymmetric and stronger for stated preferences. In other words, subjects whose choices depart more from the canonical expected utility framework seem to benefit the most from the revealed preference approach. 

In addition to the observed heterogeneity, we find that subjects often do not recognize which approach is more effective for them. In the experiment, after subjects make their choices and write their prompt, we also ask them to choose exactly one of the two AI agents to delegate decisions to. Although subjects largely choose to delegate to the agent they believe is better, they substantially overestimate the match rate difference and often misjudge which agent performs better. As a result, the proportion of subjects who delegate to the better-performing agent is not significantly higher than the 50\% random-guessing benchmark.

As a robustness check, we replicate our analysis using GPT-5.4. Our main results remain both qualitatively and quantitatively similar, suggesting that they extend beyond a single AI model. We note one interesting difference appeared in these models: when stated and revealed preference information yields conflicting predictions, Claude is more likely than GPT to follow written instructions. This is consistent with known differences in 
how these two large language models are trained \citep{bai2022constitutional, ouyang2022training}.

Taken together, our findings suggest that individuals face difficulty in articulating their preferences in economically relevant settings, and that revealed preferences can offer human principals an effective alternative that is readily implementable with a small amount of choice data. 
However, the relative effectiveness of the two approaches varies considerably across individuals, and people often fail to understand which works better for them. Thus, the success of the revealed preference approach in communicating human preferences to AI agents depends on careful implementation. 

\subsection*{Related Literature}\label{sec:related_literature}
Our paper contributes to a burgeoning literature on AI agents as economic actors.\footnote{See recent surveys, e.g., \cite{immorlica2024generative} and \cite{hadfield2025economy}.} A central concern in this literature is the potential misalignment between an AI agent’s objective and that of its human principal, for instance, when the two have different preferences  \citep{gabriel2020artificial, ji2023ai, potter2026peer, suleymanov2026revealed}. Beyond inherent preference differences, misalignment can also arise from technical limitations of AI agents, such as inaccurate inference or hallucination \citep{huang2025survey}, or from the human principal’s inability to fully articulate their preferences to the AI agent \citep{imas2025agentic}. We focus on this latter channel, which we view as the primary source of misalignment in our setting.\footnote{A growing literature supports this view, showing that LLMs exhibit consistent decision-making in choice under risk contexts \citep{chen2023emergence}. For example, \citet{kim2024learning} show that GPT’s recommendations are consistent with expected utility maximization and can be aligned with subjects’ risk aversion when provided with simulated choice data. Moreover, broader AI alignment concerns are unlikely to affect our results: our tasks involve standard choices under risk, with no safety \citep{amodei2016concrete, hendrycks2023overview} or constitutional-AI constraints \citep{bai2022constitutional} that prevent the AI agent from implementing the risk preferences stated by human principals.} 

As is well known, prompt writing is hard, and there is no reliably effective universal strategy \citep{zamfirescu2023johnny, meincke2025a, meincke2025b, meincke2025c}.
Recent evidence from  \cite{imas2025agentic} also shows substantial heterogeneity in human principals' ability to state their preferences, with these differences carrying through to economic outcomes when decisions are implemented by AI agents. This source of misalignment may become even more severe in high-dimensional decision problems, where human principals could face greater difficulty articulating their preferences precisely \citep{shahidi2025coasean, liang2026clones}. Our work shows that revealed preference information, as an alternative channel for communicating human principals' preferences, can help mitigate this form of misalignment.

Our experimental design is similar in nature to recent work on procedural rationality \citep[see, e.g.,][]{choi2006substantive, halevy2024identifying, arrieta2025procedural}, as subjects are asked to generate text-based descriptions of their preferences that may encode decision rules. The main question addressed in this line of work is whether and when preferences that are consistent with some ``logical'' decision rule\textemdash not necessarily rational in the standard utility-maximizing sense\textemdash can be identified. Our work instead focuses on the downstream task of communicating such a rule, regardless of its consistency with rationality, to an AI agent that makes decisions on a subject's behalf. 

Our paper is also related to work that uses LLMs to extract preferences from natural language for economic applications \citep{manning2025market, huang2025accelerated, li2023eliciting}. 
Rather than studying how to improve preference elicitation through natural language, we compare alternative inputs for enabling AI agents to implement human principals' preferences in a controlled and incentivized environment.\footnote{Other papers have also collected stated and revealed preferences in an unincentivized manner, such as \cite{fedyk2024ai} and \cite{lai2026users}. However, they elicit stated preferences only in the more limited form of Likert scales rather than richer free-text responses, and they do not examine LLM responses to the two information types separately. They also study different settings: the former augments LLMs with demographic information, while the latter focuses on writing assistance.} More broadly, our work contributes to the literature on human-AI interaction, particularly studies that examine human beliefs about AI capabilities and how these beliefs shape delegation decisions \citep{vafa2024large, kimbrough2026trust, he2025misperception, dreyfuss2025humanlearningai}.\footnote{For a recent survey, see \cite{jackson2025behavioral}.} 
By studying subjects' beliefs about the relative effectiveness of different information types in communicating preferences to AI, we show that misperceptions about this relative effectiveness also matter for human-AI interaction.

\section{Experimental Design}
\label{sec:design}

\paragraph{Structure of the Experiment.}
The experiment consists of three parts. Parts I and II are the main components of the experiment, in which subjects make choices under risk, write instructions for an AI agent, and decide what type of information they would like to provide to an AI agent that will later act on their behalf. 

In Part I of the experiment, subjects complete three tasks. They are informed that their decisions in each task will be used to instantiate an AI agent. This agent will then make decisions on their behalf in a new but structurally similar set of lottery problems that may be used to determine their bonus payment. In the first task, subjects choose between 13 pairs of lotteries, which are all presented in random order. These lottery problems are classified into three different categories\textemdash ``easy'',  ``hard'', and ``behavioral''\textemdash and we will describe these lottery specifics below. Subjects also have access to five summary statistics about each lottery to assist their decision-making.\footnote{These statistics are the expected value (displayed as “Average Payment”), the variance (displayed as “Payment Variability”), the minimum payment, the maximum payment, and the probability of receiving the maximum payment. See \cite{arrieta2025procedural} for a similar implementation.} We refer to the resulting AI agent endowed with this choice information as \textbf{Data-AI}.

In the second task, after completing their choices, subjects write a free-form prompt describing their preferences for choosing between lotteries, which will then be given to an AI agent. Subjects are encouraged and incentivized to provide clear, flexible guidance that can be used across different situations, but do not explicitly provide prompt examples to prevent anchoring \citep{furnham2011literature}. While writing their prompt, subjects have access to their previous choices, including the summary statistics of the lotteries. Finally, to limit the opportunity cost of time spent on the prompt, a minimum of 60 seconds is enforced on the prompt-writing screen before subjects can progress \citep{spiliopoulos2018bcd}. We refer to the resulting AI agent endowed with the prompt as \textbf{Prompt-AI}.

In the third task, subjects decide which type of information to provide to the AI agent instantiated from this task. They must select exactly one of two options: their \textit{past choice data} from task 1 or their \textit{written instructions} from task 2. We refer to this as their ``delegation decision.''

In Part II of the experiment, subjects face a new set of 13 lottery pairs designed to be structurally similar to those in Part I. Subjects' choices in these problems serve as the ground truth for evaluating AI agent predictions. However, subjects are not informed, when making these decisions, that their choices will be used to benchmark AI accuracy. This avoids subjects potentially choosing in accordance with their written prompts or past choices to maximize AI match rates as opposed to choosing what they prefer.\footnote{In both Part I and Part II, a brief brain break is implemented after every six binary lottery questions to reduce fatigue. During these breaks, subjects are asked to locate a camouflaged animal in two distinct images.}

Part III of the experiment consists of supplemental measures. We elicit subjects’ beliefs about the prediction accuracy of Prompt-AI and Data-AI by asking them to guess how many questions each AI agent predicts correctly. Subjects earn an additional 25 cents for each correct guess. We also ask subjects why they made their delegation decision. In addition, subjects complete a series of control tasks, containing a short IQ test (ICAR, \citealp{condon2014international}), risk attitude elicitation using two investment tasks \citep{gneezy1997experiment}, and measures of overconfidence \citep{moore2008trouble, chapman2023econographics}. 
 At the end of the experiment, subjects answer a brief, unincentivized survey covering self-reported comfort with AI tools, literacy, impatience, frequency of AI use, education level, and management experience, as well as a personality survey (TIPI; \citealp{gosling2003tipi}). We additionally collect demographic variables such as age and gender. See more details on these measures in the Supplemental Appendix.

\paragraph{Incentives.} The payment of subjects consists of several parts. First, subjects receive a \$5 completion payment. Second, to incentivize subjects across Parts I and II, one of four sets of 13 lottery pairs is randomly selected for a bonus payment. From the selected set, \emph{one} decision is drawn at random, with equal probability across the 13 problems, and the corresponding lottery is implemented for the bonus payment. 

Out of the four sets of 13 lottery pairs, three consist of AI predictions for the Part II lottery problems, generated using different sources from Part I: one from Data-AI based on the subject’s past choices; one from Prompt-AI based on the subject’s written prompt; and one from either, based on the subject's delegation decision. The fourth set consists of the subject’s own 13 choices from Part II. By randomly selecting one set for their bonus payment, this ensures that subjects have incentives to make genuine choices, write informative prompts, and delegate to the AI agent so that it can best act on their behalf. Finally, all Part III decisions are incentivized, except the brief summary survey at the end. 

\paragraph{Implementation.}\label{sec:implementation}
The experiment was programmed in oTree \citep{chen2016otree} and administered via Prolific across three independent sessions (February, March, and April 2026), with recruitment restricted to U.S.-based adults. 
We excluded subjects who failed a short comprehension quiz at the beginning of the experiment. A total of 296 subjects completed the experiment. For our analysis, we further excluded six subjects whose written prompts, upon close inspection, were completely uninformative about their preferences.\footnote{As an example, two of them submitted empty prompts, and one wrote ``select A''.} In addition, to ensure that prompts were written by human subjects, we also excluded 23 subjects for whom we detected that content had been pasted into their written prompts. This left us with 267 subjects in total. The experiment lasted on average 27 minutes. Subjects earned  \$12 on average, including the \$5 completion fee.

\begin{sloppypar}
All predictions made by the AI agent were generated using Claude Opus 4.5 (\texttt{claude-opus-\allowbreak4-5-20251101}), with extended thinking enabled. Each AI agent is prompted to reason through each lottery pair and submit its choices in a standardized format.  The study was approved by the Monash Human Research Ethics Committee (ID: 50731), and was pre-registered on AsPredicted (\# 268985). The full system prompts used for each AI agent, instructions, and screenshots of the interface are presented in the Supplemental Appendix.
\end{sloppypar}

\paragraph{The Lottery Choices.}\label{sec:lotteries}

Human subjects face two sets of 13 lottery pairs, one in Part I and one in Part II. Within each set, the 13 pairs span three conceptually distinct types of decisions: ``easy'', ``hard'', and ``behavioral''. Easy questions involve either first-order stochastic dominance (FOSD) or large differences in expected value between the two lotteries. Hard questions, in contrast, involve lotteries with smaller expected value differences, more outcomes on average, no obvious heuristic solution, or require reasoning based on more subtle dominance concepts such as second-order stochastic dominance (SOSD). Finally, behavioral questions are designed to test for (reverse) common ratio and (reverse) common consequence effects\textemdash two canonical environments in which violations of expected utility theory are frequently observed empirically (\citealp{blavatskyy2022experimental, blavatskyy2023common}). All lottery specifications can be found in the Supplemental Appendix.

Each part includes two easy questions and five hard questions. In Part I, we source one easy question and four hard questions from \cite{agranov2017stochastic}.\footnote{In \citet{agranov2017stochastic}, lottery outcomes are reported in tokens rather than dollar amounts. We convert those token values into dollars and present all outcomes to subjects directly in monetary terms.} The remaining easy and hard questions feature lottery pairs ordered by FOSD and SOSD, respectively. In Part II, we increase the stakes tenfold for one easy question and one hard question from Part I. For the remaining questions, we introduce slight perturbations while keeping the expected value differences within each pair largely unchanged. 

In each part, we also generate two sets of three behavioral questions using the framework of \cite{mcgranaghan2026connecting}. Specifically, for fixed prizes $H>M>0$, consider the following three binary choice questions that are parameterized by a vector $(p, r)$ where $p, r\in (0,1)$:\footnote{Here, lotteries are represented as $(x_1, p_1;\dots, x_n, p_n)$ where each outcome $x_i$ occurs with probability $p_i \geq 0$ and $\sum_{i=1}^n p_i = 1$. To simplify notation, we omit the null outcome of $0$ and use $(x, p)$ to denote the prospect $(x, p; 0, 1-p)$.}

\begin{enumerate}
    \item[(i)] $AB$ choice: choose lottery $A=(M, 1)$ or lottery $B=(H, p)$.
    \item[(ii)] $AB'$ choice: choose lottery $A=(M, 1)$ or lottery $B'=(H, pr; M, 1-r)$
    \item[(iii)] $CD$ choice: choose lottery $C=(M, r)$ or lottery $D=(H, pr)$.
\end{enumerate}
Under expected utility theory, multiplying the probabilities of non-zero outcomes by a common factor $r$, or replacing a shared consequence of a $1-r$ probability of $\$ M$ with a $1-r$ probability of $\$0$, should not change preferences. However, these transformations are routinely observed to affect choices. The common ratio effect is identified by a preference reversal between the $AB$ and $CD$ tasks: individuals prefer $A$ to $B$ but prefer $D$ to $C$. The common consequence effect is identified by an analogous preference reversal between $AB'$ and $CD$. The reverse effects correspond to the opposite choice patterns: preferring $B$ (or $B'$) to $A$ and $C$ to $D$.

For the behavioral problems in Part I, we choose $(p, r) = (4/5, 1/4)$ and $(p, r) = (1/2, 1/2)$. For the behavioral problems in Part II, we choose $(p, r) = (10/11, 11/100)$ and $(p, r) = (3/10, 1/2)$. We additionally eliminate certainty for option $A$ in Part II, using $A=(M, 0.95)$ instead. These parameterizations are chosen to generate a variety of patterns \citep{mcgranaghan2026connecting}.\footnote{In Part I, the former set of parameters is expected to generate a common ratio effect but no common consequence effect, while the latter is expected to generate strong reverse common ratio and strong reverse common consequence effects. Likewise, in Part II, the former is expected to generate strong common ratio and common consequence effects, while the latter is expected to generate only a reverse common consequence effect.}

\section{Results}
\label{sec:results}
In this section, we first compare the effectiveness of stated versus revealed preference information (Section \ref{sec:main}) and examine the sources of the performance gap (Section \ref{sec:prompt_info} ). We then examine how relative performance varies across subjects and whether they recognize which approach works better for them (Section \ref{sec:heterogeneity}). Finally, we assess the robustness of our results using GPT-5.4 (Section \ref{sec:gpt_5_4}).

\subsection{Revealed versus Stated Preferences}
\label{sec:main}

Our primary outcome of interest is the AI agent's \emph{match rate}. This is  a subject-level, out-of-sample measure, defined as the fraction of Part II decisions for which the AI agent’s prediction coincides with the subject’s actual choice.   We interpret the match rate as measuring how effectively subjects communicate their preferences to the AI agent. We find that Data-AI systematically outperforms Prompt-AI across subjects. At the aggregate level, Data-AI achieves an average match rate of 74\%, which is three percentage points higher than the 71\% achieved by Prompt-AI ($p < 0.001$). This advantage is also present across all three question categories discussed in Section~\ref{sec:lotteries}.\footnote{The average match rates for Data-AI are 99\%, 70\%, and 69\% for easy, behavioral, and hard questions, respectively. The corresponding average match rates for Prompt-AI are 95\%, 66\%, and 66\%, respectively. Paired $t$-tests reject the null hypothesis of equal match rates in all three categories ($p<0.001$, $p=0.031$, and $p=0.030$, respectively).}

To put the match rate of AI agents into perspective, we examine how Data-AI's predictive performance compares with that of a standard economic model. Given that both models use subjects' past choice information, this comparison provides a benchmark for how well Data-AI processes such information. Specifically, for each subject we estimate a constant relative risk aversion expected utility model with logit choice error over subjects' choices in Part I using maximum likelihood. We then use the estimated risk parameters to predict subjects' choices in Part II. We find that on average, the model achieves a match rate of 75\%, which is statistically indistinguishable from that of Data-AI ($p = 0.295$). At the subject-question level, Data-AI's predictions also coincide with those of expected utility 80\% of the time. Thus, Data-AI appears to effectively process the choice information it is given, as its performance is comparable to that of a standard structural model of risk preferences.\footnote{We also replicate the exercise with cumulative prospect theory (\citealp{tversky1992advances}), which accommodates the Allais paradox, and find similar results. See more details in the Supplemental Appendix.} %

We next ask the minimum number of questions that Data-AI needs to match the performance of Prompt-AI. This question is important for practical reasons. If only a few questions suffice, the revealed-preference approach can be cost-effective. By contrast, if matching the stated-preference approach requires all 13 questions, one may need to weigh the cost of eliciting many choices against the potential gain in predictive accuracy. 

To answer this question, we vary the number of Part I choices shown to Data-AI before it predicts subjects' Part II choices. 
Since questions from different categories may differ in their predictive value, we examine each category separately.\footnote{To balance the representation of the possible subsets, assignments are based on subject IDs, which were initially assigned at random. For example, when Data-AI is shown a subject's choices on two hard questions, there are ${5 \choose 2}=10$ possible subsets, which are assigned to subjects according to their ID.} As a baseline, an AI agent given no subject information achieves a match rate of 65\%. With two easy questions, it improves to 68\%, but still below Prompt-AI's match rate of 71\% ($p = 0.014$). In contrast, providing an AI with subjects' choices on either two behavioral or two hard questions is sufficient to match Prompt-AI’s average performance. Additional choices in each category yield positive but diminishing gains in the match rate. Hence, with appropriately selected questions, much of the predictive value of revealed preferences can be obtained from just a few choices, suggesting that the approach can be practical and cost-effective.\\

\textbf{Result 1:} \textit{On average, humans better communicate their preferences to AI agents using revealed rather than stated preferences. As few as two observed choices can be sufficient to match the predictive value of stated-preference information.}

\subsection{Prompt Informativeness} \label{sec:prompt_info}

Why does Data-AI do better? Our analysis suggests that the gap arises because  subjects' written prompts are not sufficiently informative. This provides empirical support for recent concerns about the difficulty humans face in communicating their preferences to  AI agents \citep{imas2025agentic, shahidi2025coasean}.

We present two pieces of evidence. First, we show that it is possible to construct prompts with predictive performance comparable to that of choice data.  
To this end, for each subject, we use Claude Opus 4.5 to generate a preference description based on all 13 of that subject's Part I choices, using the exact information and instructions given to subjects. As a descriptive exercise, we compare the most frequent words in human-written and AI-generated prompts and find 
clear differences in emphasis: %
words such as ``highest'' and ``chance'' appear five times more frequently in human-written prompts, whereas words such as ``expected'' and ``value'' appear seven times more frequently in AI-generated prompts.\footnote{See the Supplemental Appendix for word clouds generating the top 100 words that are most likely to appear in each set of prompts. The AI-generated prompts are also substantially longer, averaging 1,546 characters compared with 239 for human-written prompts.} 

We then repeat the prediction exercise using AI-generated prompts and examine match rates in Part II. The resulting AI agent achieves an average Part II match rate of 73\%, significantly higher than that of Prompt-AI ($p = 0.004$) and statistically indistinguishable from that of Data-AI ($p = 0.388$). In other words, replacing human-written prompts with AI-generated versions essentially closes the performance gap between Prompt-AI and Data-AI.

Second, we show that greater semantic similarity between human-written and AI-generated prompts is associated with greater prediction agreement and a smaller absolute match rate gap between Prompt-AI and Data-AI. To measure this textual similarity, we use SBERT \citep{reimers2019sentence}\textemdash a commonly adopted framework that maps each piece of text into a fixed-length, real-valued vector embedding by using a BERT-based language model \citep{devlin2019bert}. We then compute the cosine similarity between the human-written and AI-generated prompt embeddings, with higher values indicating greater semantic similarity.

Intuitively, predictions based on more semantically similar prompts should agree more often. Since AI-generated prompts perform comparably to Data-AI on average, we might also expect the absolute match rate gap between Prompt-AI and Data-AI to narrow as prompt similarity increases. The results in Table \ref{tab:reg_sbert_mixed_simple} are consistent with both predictions.\\

\begin{table}[t!]
\centering
\caption{Textual Similarity, AI Agreement, and Match Rate}
\label{tab:reg_sbert_mixed_simple}
\small
\scalebox{0.88}{
\begin{tabular}{lcc}
\toprule
 & (1) & (2) \\[2mm]
 & Agreement Rate
 & Absolute Match Rate Gap \\
\cmidrule(lr){2-2} \cmidrule(lr){3-3}
 & \shortstack{AI-generated vs. \\ Human-written prompt}
 & \shortstack{Data-AI vs. \\ Prompt-AI} \\
\midrule
SBERT Cosine Similarity & 0.203** & -0.083* \\
          & (0.082) & (0.043) \\[4pt]
Constant & 0.613 & 0.160 \\
         & (0.054) & (0.029) \\[4pt]
\midrule
Observations & 267 & 267 \\
$R^2$ & 0.022 & 0.013 \\
\bottomrule
\end{tabular}
}
\begin{flushleft}
\footnotesize
\textit{Note:} Both dependent variables are measured in Part II. Robust standard errors are reported in parentheses. $^{*}p<0.10$, $^{**}p<0.05$, $^{***}p<0.01$.
\end{flushleft}
\end{table}

\textbf{Result 2:} \textit{The performance gap between Prompt-AI and Data-AI largely reflects human subjects’ difficulty in articulating their preferences in text.} %

\subsection{Heterogeneity in Preference Communication}\label{sec:heterogeneity}

\begin{figure}[t]
    \centering
    \caption{Comparison of Prompt-AI and Data-AI match rates.} 
    \includegraphics[width=0.55\textwidth]{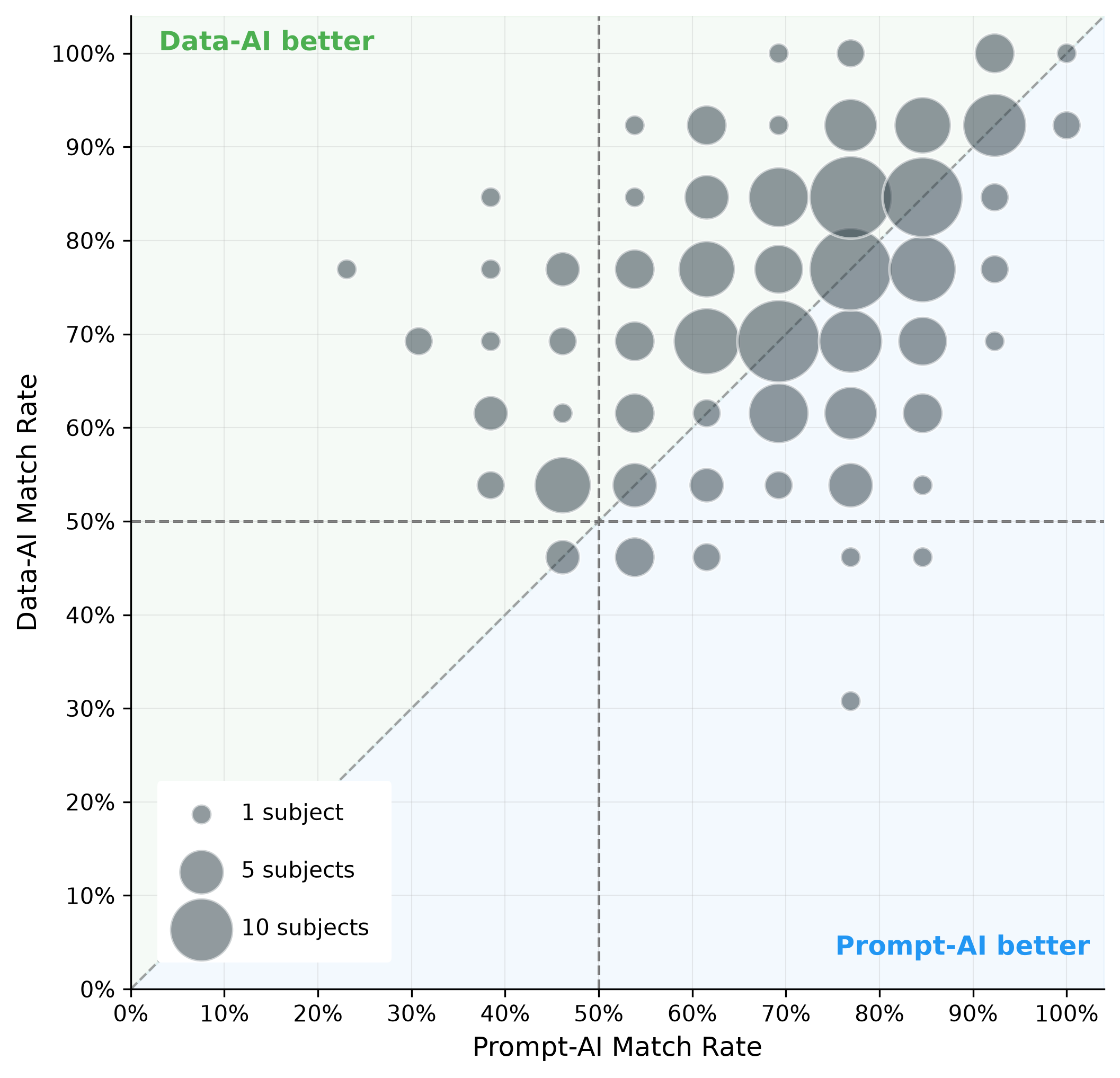}
   \captionnote{\textit{Note:}
Each bubble plots a subject's Data-AI match rate against their Prompt-AI match rate, with bubble area proportional to the number of subjects at that point. Subjects above the 45-degree line are matched better by Data-AI; subjects below it are matched better by Prompt-AI. The dashed grey lines at 50\% mark the expected match rate of random guessing on each axis.}
    \label{fig:promptvsdata}
\end{figure}

We now examine how the relative effectiveness of stated and revealed preference information varies across subjects. Figure~\ref{fig:promptvsdata} presents a scatter plot of the mean match rate of Data-AI against that of Prompt-AI at the subject level, with larger points indicating a higher number of subjects. 
We find substantial heterogeneity. For about half of subjects (46\%), Data-AI outperforms Prompt-AI, whereas the reverse holds for only 28\% of subjects. For the remaining 27\%, the two agents achieve identical match rates. Nevertheless, for most subjects (86\%), both agents outperform the 50\% random-guessing benchmark.

\begin{table}[t!]
\centering
\begin{threeparttable}
\caption{Match Rates and Behavioral Effects}
\label{tab:reg_match_rate_combined}
\begin{tabular*}{0.98\textwidth}{@{\extracolsep{\fill}}lcccc}
\toprule
 & \multicolumn{4}{c}{Match rate} \\
\cmidrule(lr){2-5}
 & (1) & (2) & (3) & (4) \\
\midrule
Data-AI (vs Prompt-AI) & 0.034*** & 0.034*** & 0.010 & 0.010 \\
          & (0.009) & (0.009) & (0.013) & (0.013) \\
No.\ Behavioral Effects &  & -0.033*** & -0.041*** & -0.042*** \\
          &  & (0.006) & (0.007) & (0.008) \\
Data-AI $\times$ No.\ Behavioral Effects &  &  & 0.015** & 0.015** \\
          &  &  & (0.007) & (0.007) \\
          Constant & 0.706 & 0.759 & 0.771 & 0.773 \\
         & (0.009) & (0.013) & (0.014) & (0.014) \\[4pt]
\midrule
Controls & No & No & No & Yes \\[2pt]
Observations & 532 & 532 & 532 & 532 \\
Subjects     & 266 & 266 & 266 & 266 \\
$R^2$        & 0.014 & 0.091 & 0.095 & 0.153 \\
\bottomrule
\end{tabular*}
\begin{tablenotes}[flushleft]
\footnotesize
\item \textit{Note:} Standard errors clustered by subject in parentheses. Each subject contributes two match rate observations (Prompt-AI and Data-AI). Data-AI is an indicator variable for whether the match rate is for Data-AI (versus Prompt-AI). No.\ Behavioral Effects is a count variable ranging from 0 to 4. Column (4) adds controls collected in Part III.  $^{*}p<0.10$, $^{**}p<0.05$, $^{***}p<0.01$. See the full regression for Column (4) in the Supplemental Appendix.
\end{tablenotes}
\end{threeparttable}
\end{table}

To better understand this heterogeneity, we first examine how individual characteristics relate to the performance gap between the two approaches. These characteristics include IQ, risk attitudes, and personality traits, which we measure using control tasks and surveys in Part III of the experiment. We find that the performance gap between Data-AI and Prompt-AI is smaller for subjects who are more comfortable with AI and more patient, possibly reflecting greater ability or willingness to formulate informative prompts. In addition, greater compound risk aversion is associated with a larger performance gap in favor of Data-AI. This may reflect broader departures from expected utility that are more readily captured by observed choices than by subjects’ written descriptions of their preferences.\footnote{See the Supplemental Appendix for more details.}

Next, we classify subjects based on their tendency to exhibit behavioral effects in Part I, measured by the number of times they exhibit the (reverse) common ratio or (reverse) common consequence effects. Since there are two sets of behavioral questions, and each set allows us to observe one (reverse) common ratio effect and one (reverse) common consequence effect, the number of behavioral effects ranges from zero to four. Overall, 25\% of subjects exhibited none of these effects, 52\% exhibited one or two, and 23\% exhibited three or four.

Table \ref{tab:reg_match_rate_combined} relates an AI agent’s mean match rate in Part II to a Data-AI indicator, the number of behavioral effects, and their interaction. 
Consistent with our earlier findings, Column (1) shows that on average, Data-AI outperforms Prompt-AI by about three percentage points. Columns (2)-(4) show that across both approaches, each additional behavioral effect is associated with a three- to four-percentage-point decrease in match rate. In other words, subjects who exhibit more behavioral effects seem to be less effective at communicating their preferences through either approach.

Interestingly, the decline in match rates associated with behavioral effects is more pronounced for Prompt-AI than for Data-AI. As seen in Column (3), among subjects who exhibit no behavioral effects, there is no significant performance gap. Each additional behavioral effect is associated with a 4.1-percentage-point decrease in Prompt-AI’s match rate, compared with a 2.6-percentage-point decrease for Data-AI. Thus, among those who exhibit all four behavioral effects, Data-AI's advantage is about six percentage points larger.
Column (4) demonstrates that these relationships remain robust after controlling for demographic and survey characteristics. 

One possible interpretation is that people who exhibit more behavioral effects have higher-dimensional preferences that are harder to articulate.\footnote{For example, consider two subjects, one whose preferences are consistent with EU, and another whose preferences are consistent with a non-EU model, such as prospect theory. The former need only provide enough information to pin down a Bernoulli utility function, while the latter must also specify a probability weighting function.} Their choices, however, may naturally integrate these dimensions without requiring subjects to articulate them separately. As a result, the advantage of revealed-preference information may be more pronounced for these subjects.

\begin{figure}[t!]
    \centering
    \caption{Delegation Choice: Realized Accuracy and Beliefs}
    \includegraphics[width=\textwidth]{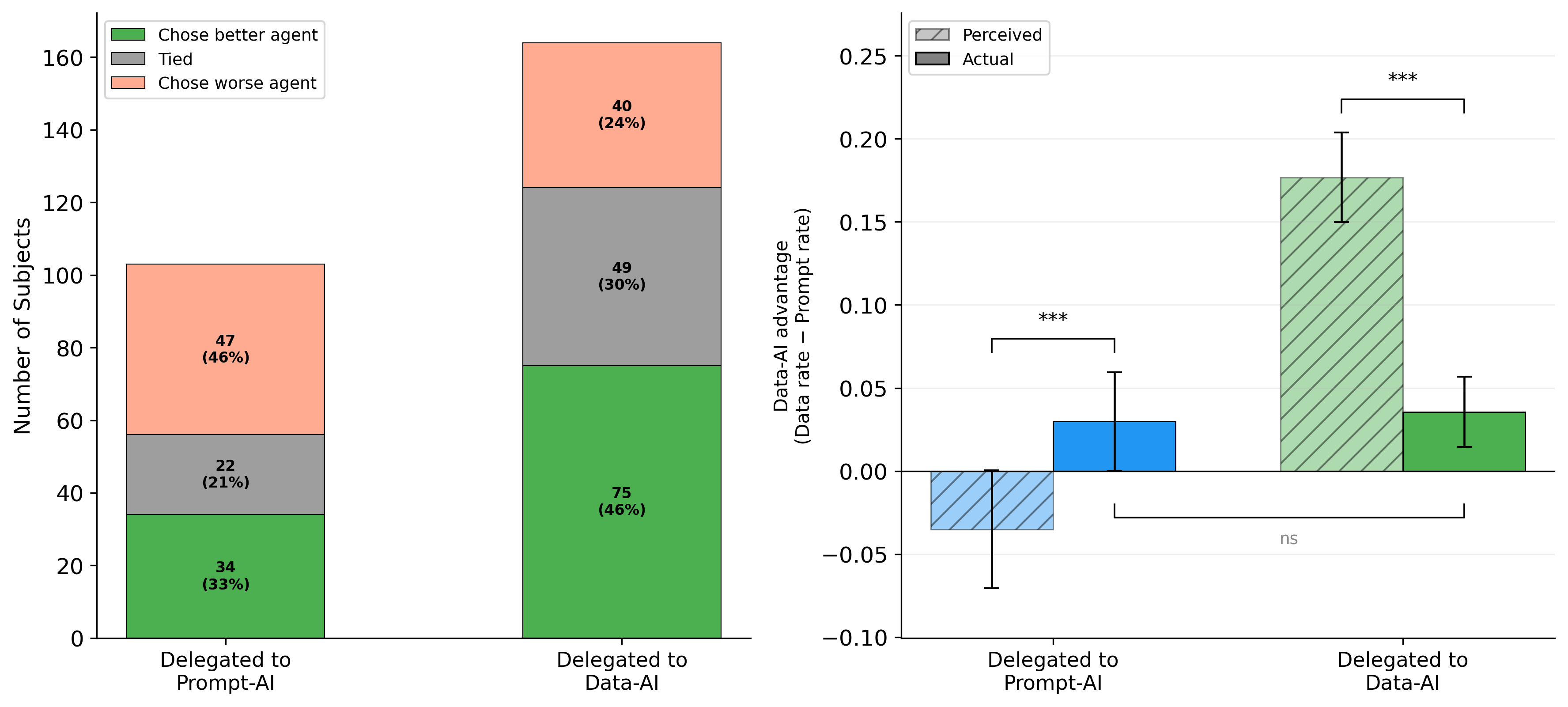}
    \label{fig:delegation}
    \captionnote{\textit{Note:} \textit{Left}: counts (within-group percentages) choosing the better agent (green), tied agents (grey), or the worse agent (salmon). \textit{Right}: mean Data-AI–Prompt-AI match-rate differences, actual (solid) and perceived (hatched), with 95\% confidence intervals. Upper brackets: paired t-tests of perceived versus actual differences within delegation groups. Lower bracket: two-sample t-test of actual differences across groups. $^{*}p<0.10$, $^{**}p<0.05$, $^{***}p<0.01$.}
\end{figure}

\paragraph{Delegation and Beliefs.}\label{sec:delegation}
Finally, we examine whether subjects recognize which approach better communicates their preferences and delegate appropriately. Our delegation data suggest that they often do not. The left panel of Figure \ref{fig:delegation} reports the number of subjects by their delegation decision and classifies each group according to whether the chosen agent actually achieves a higher, equal, or lower match rate than the alternative. Excluding subjects for whom the two agents are equally accurate, 56\% delegate to the better-performing agent \textemdash a proportion that is not significantly different from the 50\% random-guessing benchmark ($p=0.133$).

This delegation failure appears to reflect subjects' inaccurate beliefs about the relative performance of the two AI agents. We next examine their perceptions of each AI agent's performance, elicited in Part III of the experiment. The right panel of Figure \ref{fig:delegation} shows the realized and perceived differences in match rates between Data-AI and Prompt-AI, conditional on the subjects' delegation decision. Overall, 82\% of subjects delegate to the AI agent they expect to perform weakly better, suggesting that their delegation decisions align with their beliefs about relative AI agent performance. However, among subjects who delegate to Prompt-AI, on average, they believe that Prompt-AI outperforms Data-AI, when in fact the opposite is true.
Among subjects who delegate to Data-AI, they are correct about the direction of relative performance on average, but substantially overestimate its magnitude. Thus, while delegation decisions largely reflect subjects' perceived performance of AI agents, these perceptions are often inaccurate in either direction or magnitude.\\ 

\textbf{Result 3:} \textit{About three-quarters of subjects communicate their preferences to an AI agent at least as effectively through revealed preferences as through stated preferences. The relative advantage of revealed-preference information is particularly pronounced among subjects who exhibit more behavioral effects. Subjects, however, often do not recognize which approach is more effective for them, leading many to delegate to the worse-performing agent.}

\subsection{Robustness: GPT-5.4}\label{sec:gpt_5_4}

As a robustness check, we now examine whether our results depend on the choice of Anthropic’s Claude Opus 4.5 as the AI agent. In the Supplemental Appendix, we replicate our analysis by replacing Claude Opus 4.5 with GPT-5.4 as the underlying AI model.\footnote{Note that in the experiment, subjects were explicitly informed that Claude was the AI model used to determine their payment. This may have affected the instructions they wrote, their delegation decisions, and their perceptions of AI performance.}
Our results remain both qualitatively and quantitatively similar when using GPT, suggesting that they extend beyond a single AI model. 

We note two interesting differences. First, the AI agent using GPT-generated prompts has a mean match rate of 71\%, which, unlike Claude, no longer outperforms Prompt-AI. This appears to be driven by the lower quality of the GPT-generated prompts rather than by the AI model itself. For example, when GPT is instead given the Claude-generated prompts, its mean match rate rises to 74\%, on par with Claude’s performance using the same prompts. 

Second, different AI models seem to place different weights on stated versus revealed preference information when the two conflict. Consider subject-question pairs in which Data-AI and Prompt-AI predict different choices. An agent given \textit{both} types of information must then agree with one of these two predictions. We find that such an AI agent is significantly more likely to follow Prompt-AI when the underlying model is Claude rather than GPT.\footnote{For Claude, such conflicts arise in 25\% of subject-question pairs; for GPT, the corresponding rate is 28\%. Among each model's own conflicting pairs, Claude follows Prompt-AI 67\% of the time, while GPT follows it only 49\% of the time.} This pattern is consistent with known differences in the post-training methods used by these two AI models: Anthropic’s emphasis on constitutional-style alignment \citep{bai2022constitutional} explicitly prioritizes stated goals from a human user, whereas OpenAI uses reinforcement learning from human feedback \citep{ouyang2022training}, which does not prioritize stated preferences in the same manner. 

\section{Conclusion}\label{sec:conclusion}
This paper shows that human principals communicate their preferences more effectively to AI through a revealed-preference approach than through a stated-preference approach. Our analyses suggest that this advantage reflects subjects' difficulties in articulating preferences in text and is greatest among those who exhibit more behavioral effects. Furthermore, subjects often do not recognize which approach is better for them, leading many to delegate to a less accurate agent.

Our framework and results suggest several directions for future research. It would be interesting to consider settings in which stated and revealed preferences may diverge, for example due to present bias  \citep{o2015present} or social desirability bias  \citep{norwood2011social}. Such divergence complicates the evaluation of AI agents, as the human principal's true preferences may be unclear \citep{kleinberg2024inversion}. It would also be useful to examine higher-dimensional choice environments, where AI agents may struggle to implement given preferences \citep{liang2026clones}. Exploring these directions could provide further understanding of how different environments shape the relative effectiveness of stated versus revealed preferences in guiding AI agents.

\newpage
\singlespacing

\end{document}